\documentclass[12pt,preprint]{aastex}

\slugcomment{}

\shorttitle{Broad Line Region in NGC 3227}
\shortauthors{Devereux et al.}

\begin{document}


\title{The Size, Structure and Ionization of the Broad Line Region in NGC 3227}


\author{Nick Devereux}
\affil{Department of Physics, Embry-Riddle Aeronautical University,
    Prescott, AZ 86301}
\email{devereux@erau.edu}



\begin{abstract}

{\it Hubble Space Telescope} spectroscopy of the Seyfert 1.5 galaxy, NGC 3227, confirms previous reports that the broad H${\alpha}$ emission line flux is time variable, decreasing by a modest ${\sim}$ 11\% between 1999 and 2000 in response to a corresponding  ${\sim}$ 37\% decrease in the underlying continuum. Modeling the gas distribution responsible for the broad H${\alpha}$, H${\beta}$ and H${\gamma}$ emission lines favors a spherically symmetric inflow as opposed to a thin disk. 
Adopting a central black hole mass of 7.6 ${\times}$ 10$^{6}$ M${_{\sun}}$, determined from prior reverberation mapping, leads to the 
following dimensions for the size of the region emitting the broad H${\alpha}$ line; an outer radius ${\sim}$ 90 l.d and an inner radius ${\sim}$ 3 l.d. Thus, the previously determined reverberation size for the broad line region (BLR) consistently coincides with the inner radius of a much {\it larger} volume of ionized gas. However, the {\it perceived} size of the BLR is an illusion, a consequence
of the fact that the emitting region is ionization bounded at the outer radius and diminished by Doppler broadening at the inner radius. The actual dimensions of the inflow remain to be determined. Nevertheless, the steady state mass inflow rate is estimated to be ${\sim}10^{-2} ~{\rm M_\odot~yr^{-1}}$ which is sufficient to explain the X-ray luminosity of the AGN in terms of radiatively inefficient accretion.  Collectively, the results challenge many preconceived notions concerning the nature of BLRs in active galactic nuclei.

 \end{abstract}


\keywords{galaxies: Seyfert, galaxies: individual (NGC 3227), quasars: emission lines}



\section{Introduction}

NGC 3227 is an early type SAB(s)a spiral galaxy located at a distance of 20.8 Mpc \citep{Tul88}. The galaxy harbors a Seyfert 1.5 nucleus  \citep{ Ho97} whose broad Balmer emission lines have long been known to be time variable. Since the broad line region (BLR) is unresolved, the spatial distribution of emission cannot be directly measured.
However, reverberation mapping, which refers to measuring the time delay in the response of broad emission lines to time variable illumination from the central continuum source \citep{Pet93}, has yielded a variety of estimates for the size of the broad line region \citep{Pet85, Win95, Pet04, Kas05, Den09, Den10}.
Even though the broad line emission emanates from a finite volume, reverberation mapping yields only a single size. Despite numerous attempts to model the reverberation phenomenon \citep[e.g.,][and references therein]{Ede88, Rob90, Wel91, Hor04, Kor04, Pan11}, the question remains {\it what exactly does the reverberation size refer to;  the inner radius of the BLR, the outer radius or a luminosity weighted radius?} 

Emission line profile fitting is a complementary technique used to constrain the physical dimensions of the BLR in low luminosity AGNs (LLAGNs) with known central masses. Gravity dominates the kinematics of the gas because LLAGNs radiate well below the Eddington limit with insufficient radiation pressure to drive an outflow. 
Consequently, knowing the relationship between velocity and radius allows the broad emission line profiles to be modeled revealing the shape and size of the BLR. 
This technique has led to large BLR size discrepancies for the LLAGNs M81, NGC 3998 and NGC 4203 \citep{Dev07, Dev11a, Dev11b}. Specifically, the dimensions of the BLR deduced from emission line profile fitting are much larger than expected based on the reverberation size--UV luminosity correlation of \cite{Kas05}. However, \cite{Kas05} note that the correlation appears to break down for AGNs with low UV luminosities, comparable to those measured for M81, NGC 3998 and NGC 4203, so the comparison may not be meaningful. On the other hand, the reverberation size has actually been measured for the LLAGN in NGC 3227, most recently by \cite{Den10}.
The principle aim of this paper, therefore, is to make the first direct comparison of the reverberation size with the dimensions of the BLR in NGC 3227 deduced from emission line profile fitting. 
The larger context for this investigation is to understand why it is that BH masses estimated using the reverberation radius require a considerable factor of ${\sim}$ 5.5 
correction in order to place them on the M$_{\bullet}$--${\sigma}_* $ relation defined by BH masses measured directly using gas and stellar kinematics \citep{Onk04}.

The mass of the BH in NGC 3227 has been measured using three different methods; stellar kinematics \citep{Dav06}, gas kinematics \citep{Hic08} and most recently via reverberation mapping \citep{Den10}, collectively
yielding mass estimates in the range (0.7 -- 2) ${\times}$ 10${^7}$ M$_{\odot}$. When combined with the X-ray luminosity the range of BH masses indicate that NGC 3227 is radiating at ${\leq}$0.4\% of the Eddington luminosity limit \citep{Win11, Xu11, Vas09}.

The layout of the paper is as follows. In Section 3, the broad emission lines seen in NGC 3227 are evaluated in the context of inflow and accretion disk models. Some physical properties of the BLR are presented in Section 4 including the size, structure and source of ionization. Conclusions follow
in Section 5. We begin, however, with Section 2 and a review of the emission lines observed in the nucleus of NGC 3227.

\section{NGC 3227 Emission Lines }


NGC 3227 has been observed with STIS twice; once in 1999 with the G750M grating and again in 2000 with the G750L, G430L, G230L and G140L gratings. Table 1 summarizes the observations and the archival spectra are presented in Figure 1. 
STIS spectra showing the UV--visible emission have been presented previously;
\cite{Cre01} based on observations obtained under PID 8479 and the G750M spectrum obtained under PID 7403, by \cite{Wal08}. All spectra shown in the following refer to a 7 pixel wide extraction along the slit direction and centered on the nucleus. Each extraction samples ${\ge}$ 80\% of the encircled energy for an unresolved point source \citep{Pro10}. It is quite obvious 
from the G750M and G750L spectra that the continuum decreased in brightness by ${\sim}$ 37\% between 1999 and 2000. 
Collectively, the G750M, G750L and G430L spectra resolve the H${\alpha}$, H${\beta}$ and H${\gamma}$ lines, a more detailed description of which follows in the next section.

\subsubsection{Broad Balmer line Emission }

Figure 2 illustrates that H${\alpha}$ emission line profile is single peaked and both of the [N II] vacuum wavelength 6549.85 {\AA} and 6585.28 {\AA} emission lines can be seen in the high resolution G750M spectrum facilitating their ultimate subtraction to reveal an unadulterated broad H${\alpha}$ emission line. 
That the tips of the [N II] lines are clearly visible provides a useful constraint on their flux. Rather than constraining the width of the [N II] lines to be the same as the [S II] lines \citep[e.g.,][]{NS03,Wal08}, the width and the flux of the brightest [N II] line, and by association the fainter [N II] line, were determined empirically by selecting the values that resulted in the {\it cleanest} subtraction. Persistent irregularities
in the broad H${\alpha}$ emission line profile may be real as they do not coincide
with the central wavelengths of the subtracted [N II] lines as illustrated in Figure 2. Otherwise, the broad H${\alpha}$ emission line profile is symmetric about the ${\lambda}$6588{\AA} wavelength expected for a systemic velocity of 1126 km/s deduced from the peak of the brightest [N II] emission line. 
Thus, there is no apparent redshift between the broad H${\alpha}$ emission line and the systemic redshift of the host galaxy. 
The ${\bf STSDAS}$ contributed task ${\bf specfit}$ was used to model and subtract the forbidden lines and the results are reported
in Table 2 along with the fluxes for the two [S II] vacuum wavelength ${\lambda}$6718.29 and ${\lambda}$6732.67 {\AA} emission lines
and the two [O I] vacuum wavelength ${\lambda}$6302.04 and ${\lambda}$6365.53 {\AA} emission lines.

 It is also customary to subtract a narrow component of the H${\alpha}$ line. This practice arose because the relatively large, 1 - 2 arc sec wide, slits employed for ground based spectroscopic observations \citep[e.g.,][]{Ho95}  inevitably included ``extended" narrow line gas leading to a narrow component of H${\alpha}$ emission centered at the systemic velocity of the galaxy. However, the STIS spectra examined in this paper were obtained with a much smaller 0.2 arc sec slit (Table 1) and subsequent modeling shows that the H${\alpha}$ emission originates from an even smaller region, less than 5 ${\times}$ 10$^{-4}$ arc sec in radius ({\S 4.2, 4.3). Consequently, there is no evidence for an ``extended" narrow line region of H emitting gas. 
Besides, estimating the flux of the narrow H${\alpha}$ line is highly subjective. The model employed by \cite{Wal08} for NGC 3227 has the narrow H${\alpha}$ brighter than the brightest [N II] line whereas the narrow H${\alpha}$ line is typically {\it observed} to be 3 to 5 times fainter than the brightest [N II] line in STIS spectra of AGNs for which the lines can be clearly seen \citep[e.g.,][]{NS03}. 
Furthermore, it can be demonstrated empirically that subtracting a narrow H${\alpha}$ component with less than half the flux of the brightest [N II] line has little consequence on the 
appearance of the bright central spike in the G750M spectrum. Also, in the specific case of NGC 3227, presumably 
the H${\beta}$ and H${\gamma}$ profiles have central spikes as well, but there is no way to see them and subtract them with the low resolution spectra that are currently available. Under these circumstances it would be inappropriate to subtract a narrow component from the H${\alpha}$ line and not the other Balmer lines as well.
Another important consideration to bear in mind is that a bright central spike is a feature of an inflow just as double peaks are a feature of a disk \citep[e.g.,][]{Che89} and to subtract the bright spike from the H${\alpha}$ spectrum may inadvertently remove an important piece of evidence in deciphering the origin of the broad emission lines. 
Given the above considerations, it would seem prudent, albeit novel, to not subtract a narrow H line from any of the Balmer lines seen in NGC 3227, which is the approach adopted here.

The forbidden line model produced using the G750M spectrum is also subtracted from the lower resolution G750L spectrum,
motivated by the observation that the flux in the two [S II] lines did not change between 1999 and 2000, with the result that the broad H${\alpha}$ emission
line flux in the G750L spectrum is ${\sim}$ 11\% smaller than measured in the G750M spectrum (see Table 2), presumably
a consequence of the aforementioned decrease in the underlying continuum.   
Figures 3 and 4 illustrate the consequence of subtracting the forbidden [O III] emission lines from the H${\beta}$ and H${\gamma}$ lines following the procedure described in \cite{Dev11b}. 
Emission line fluxes are reported in Table 3 for all the lines that can be reliably resolved and measured in the G430L spectrum  including H${\beta}$, H${\gamma}$, the two [O III] vacuum wavelength ${\lambda}$4960.30 {\AA} and ${\lambda}$5008.24 {\AA} emission lines.  
Upper limits are provided for the vacuum wavelength ${\lambda}$4364.44 {\AA}  [O III] emission line and the [O II] ${\lambda\lambda}$3727.09, 3729.88 doublet. The latter is so faint 
that the [O III] ${\lambda}$5008.24 /[O II]  ${\lambda\lambda}$3727.09, 3729.88 emission line ratio measured with STIS
locate NGC 3227 outside the boundaries of the \citet[][their Fig. 5]{Kew06} diagnostic diagram. Consequently, the STIS spectra render the AGN in NGC 3227 unclassifiable.\footnotemark \footnotetext{ The complementary [O I] ${\lambda}$6302.04/H${\alpha}$ emission line ratio can not be quantified as no estimate is provided for the narrow component of the H${\alpha}$ emission line.}

\subsubsection{Balmer Decrements}

The observed Balmer decrements, 
H${\alpha}$/H${\beta}$ = 3.96 ${\pm}$ 0.01, slightly lower than the value of ${\sim}$5 reported previously using ground based telescopes \citep{Win95, Coh83},
and H${\beta}$/H${\gamma}$ = 3.15 ${\pm}$ 0.08, measured for the first time with STIS, are both significantly different from the Case B values, 2.75 and 2.1, respectively, in the sense that
the observed values are systematically 44\% and 50\% higher, respectively. Deviations from recombination theory have been noted for other AGNs \citep[e.g.,][]{Dev11a,Dev11b,Sto97,Bow96,Fil84} and have been attributed to collisional excitation in gas of high density and temperature rather than dust extinction. Indeed, the striking similarity between the H${\alpha}$, H${\beta}$ and H${\gamma}$ lines illustrated in Figure 5, when they are normalized to their respective peak intensities and the wavelength scales converted 
to velocity using the non-relativistic Doppler equation, suggests that dust extinction to the BLR in NGC 3227 is negligible.

\section{BROAD LINE REGION MODELS}

NGC 3227 is unable to sustain a radiatively driven outflow \citep[e.g.,][]{Fab06,Kin03,Mur97,Shl85} because the AGN radiates at ${\leq}$ 4 ${\times}$ 10${^{-3}}$ the Eddington luminosity limit \citep{Win11, Xu11, Vas09}. However, single peak broad Balmer emission lines, similar to those observed in NGC 3227, can be produced by an inflow \citep[e.g.,][]{Dev11a} or 
an accretion disk \citep[e.g.,][]{Era01} and both models are tested in the following. 

\subsection{Central Stellar Mass}

In an inflow or disk model the velocity law is determined by the central mass distribution, M(r), which can be modeled as a point mass, ${M_{\bullet}}$, representing the
BH, embedded in the center of an extended star cluster. The central stellar mass may be important given the relatively low mass estimated for the BH in NGC 3227. \cite{Hic08} conveniently parameterize the near-infrared, 1.65${\mu}$, $H$-band surface brightness profile of NGC 3227 in terms of a S\'ersic function which may be integrated to determine the line-of-sight, or projected, luminosity in a variety of synthetic apertures. Such projected luminosities can be converted into central masses using the mass-to-projected-light ratio provided in \cite{Dev87}. Following this methodology leads to the values presented in Table 4
which, when compared to estimates of the central stellar mass density presented previously by \citet[][ their Figure 15]{Dav06}, reveals a very large discrepancy. For example, the first row in Table 4 corresponds to an angular radius of ${\sim}$ 0.01{\arcsec} inside which the stellar mass density is estimated to be ${\sim}$ 1.3 ${\times}$ 10${^{6}}$M$_\odot$/pc$^3$. This is  ${\sim}$ 15,000 times smaller than the value estimated by \cite{Dav06} inside that same radius. The reason for the difference has not been determined but the impact of the higher stellar mass density on the gas kinematics is substantial, producing higher gas velocities further from the central BH, and altering the shape
of the model broad line profile compared to one calculated using the smaller values. Given this difference, the smaller values for the central stellar mass density, presented in Table 4, are adopted in the following analysis.

\subsection{Inflow and Relativistic Accretion Disk Models}

The spherically symmetric inflow model has been described most recently by \cite{Dev11a} and simulates a steady-state free-fall of ${\sim}$ 10${^4}$ {\it discrete} particles. 
The velocity law and mass conservation determine the number density of inflowing particles to be ${n(r) \propto r^{-3/2}}$. Consequently, there are only two free parameters; the inner and outer radii of the inflow, ${r_i}$ and ${r_o}$, respectively. A cursory exploration of the parameter space revealed that an
outer radius, ${r_{o}}$ = 1.3 ${\times}$ 10${^{5}}r_g$, and an inner radius,  ${r_{i}}$ = 9.3 ${\times}$ 10${^{3}}r_g$, where $r_g$ = ${GM_{\bullet}/c^2}$, the gravitational radius, produced a reasonable representation of the broad H${\alpha}$, H${\beta}$ and H${\gamma}$ lines illustrated in Figure 5. 

The relativistic accretion disk model, presented by \cite{Che89}, has been described most recently by \cite{Dev11b}. 
Past experience with this model has revealed that a single peak broad line profile requires a large ratio of outer radius to inner radius, sufficient inclination to produce a symmetrical profile shape, and 
adequate velocity dispersion to merge the two peaks, a feature of disk models, into a single peak. 
A cursory search of parameter space yielded a reasonable representation of the Balmer lines illustrated in Figure 5 with a dimensionless outer radius,
${\xi}_{o}$ = 3.5 ${\times}$ 10${^{5}}r_g$, an inner radius, ${\xi}_{i}$ =  2.5 ${\times}$ 10${^{3}}r_g$, an inclination angle measured from the disk normal to the line
of sight, $i$ = 50$^{\circ}$, an emissivity law of the form $r^{-q}$, where $q$ = 2 and a velocity dispersion for the gas, ${\sigma}$ = 300 km/s. 
Interestingly, Figure 5 illustrates that the  broad Balmer lines observed with the low resolution gratings can be represented equally well by an inflow or an accretion disk; the reduced ${\chi}^2$ ${\sim}$ 8 for both models.
The effective velocity resolution of the model profiles is  ${\sim}$ 230 km/s, similar to that of the G750L grating and about twice that of the G430L grating. 
In either case the residuals between the observed profiles and the model are ${\leq}$ 10\% except for the fainter and consequently noisier H${\gamma}$ line profile.
Both the disk and inflow models have a similar shortfall in that they both predict slightly less emission on the blue side of the profile than is observed.

\subsection{The Importance of Spectral Resolution}

Figure 6 shows the broad H${\alpha}$ line measured with the higher resolution G750M grating which reveals a distinctly triangular profile that is easily modeled as an inflow
when the model spectrum is binned to ${\sim}$ 27 km/s, corresponding to the velocity resolution of the G750M grating.
On the other hand, the broad H${\alpha}$ line measured with the higher resolution G750M grating is impossible to mimic with an accretion disk model which is expected to resolve two peaks at this higher spectral finesse.\footnotemark \footnotetext { The broadening
parameter in the \cite{Che89} disk model is not intended to mimic instrumental resolution. Consequently, the model disk spectrum shown in Figure 6 was generated using a non-relativistic
Monte Carlo simulation employing a total of ${\sim}$10$^{4}$ points. The resulting velocity histogram is binned by 25 km/s and subsequently smoothed using a 40 bin moving
average.}
Focusing, therefore, on the inflow model, the reduced Chi-squared statistic was computed for 32 values of the outer radius and 18 values of the inner radius in order to constrain the uncertainty on these two parameters. 
Chi-squared minimum, ${\sim}$ 6, coincided with ${r_{o}}$ = (2 ${\pm}$ 1) ${\times}$ 10${^{5}}r_g$, and ${r_{i}}$ = (7 ${\pm}$ 1) ${\times}$ 10${^{3}}r_g$ which are the parameters adopted to produce the model inflow spectrum shown in Figure 6.  

An advantage of modeling the inflow using a Monte Carlo simulation is that it reproduces the small-scale structure, seen in broad emission line profiles, caused by random clumping 
of particles in radial velocity space. Historically, the widely quoted {\it optically thick broad line cloud} model found early support by accounting for such irregularities.
That model remains popular to this day, even though the origin and confinement of such {\it clouds} is a major unsolved problem \citep{Mat87}. With hindsight, it may be more realistic to identify the {\it clouds} with {\it filaments}, motivated by the observation
that ionized, density bounded gas filaments are commonly seen in the nuclei of active galaxies \citep{Sto09} including the Galactic center \citep{Lac91}. Such filaments have approximately the same gas density and hence the same emissivity regardless of their location with respect to the central AGN, they are optically thin, emit isotropically and have a very low volume filling factor. Consequently, the ionized H gas does not constitute a fluid and is therefore not a Bondi flow \citep{Bon52}.
As noted by \cite{Cap81}, structure in broad line profiles is revealed depending
on the number of filaments and the spectral resolution of the detector. The irregularities clearly seen in the broad H${\alpha}$ emission line of NGC 3227 observed with the G750M grating 
are mimicked by the inflow model and suggest that the inflow is illuminated by the collective emission of ${\sim}$ 10${^4}$ ionized gas {\it filaments}. 




\section{Discussion}

The exquisite spectra obtained for NGC 3227 with the ${\it HST}$ reveal an important observational phenomenon; that the shape of its bright and broad H${\alpha}$ line changes with the resolution of the grating employed to observe it, as expected if the 
emission line is produced by thousands of discrete emitters as explained previously by \cite{Cap81}. Furthermore, the triangular shape observed for the line using the high resolution G750M grating is inconsistent with the profile shape expected
for an accretion disk. Whereas it is already known that the velocity field of the narrow line gas in NGC 3227 is chaotic \cite[][and references therein]{Wal08}, the symmetry of the broad H${\alpha}$ line suggests a rather ordered velocity field, demonstrably consistent with an inflow, the size, structure and ionization of which is explored further in the following.

\subsection{Broad Line Region Ionization}

The ionizing continuum generated by the AGN in NGC 3227 may be represented, in the year 2000, as a power law
with an optical to X-ray spectral index ${\alpha}$ = 1.17 \citep{Vas09} and a normalization provided by the extinction corrected UV continuum described in Section 4.2. 
Integrating the spectral energy distribution from 13.6 eV to 100 keV using the method described previously
in \cite{Dev11a} predicts 6.1${\times} {10^{52}}$ ionizing ph/s. For comparison, the broad  H${\alpha}$ emission line flux measured in the year 2000 (Table 2) corresponds to an H${\alpha}$ luminosity,  $L(H{{\alpha}}$) = 2.4${\times}$10$^7$ L${_{\sun}}$, which, assuming 45\% of the ionizing photons are converted into H${\alpha}$ photons (Case B recombination at a temperature of 10$^4$ K), requires 6.8${\times}$${10^{52}}$ ionizing ph/s using,

\begin{equation}
N_{ion} = L(H{_{\alpha}}){\alpha_{B}}/{\alpha^{eff}_{H\alpha}} h \nu_{H\alpha} 
\end{equation}

where ${\alpha^{eff}_{H\alpha}}$ = 1.16 x 10$^{-13}$ cm${^{3}}$ s${^{-1}}$ is the effective recombination coefficient  and 
${\alpha_{B}}$ = 2.59 x 10$^{-13}$ cm${^{3}}$ s${^{-1}}$ is the total Case B recombination coefficient. 

The comparison shows that the number of ionizing photons required to excite the broad H${\alpha}$ emission line measured in the year 2000 coincides with the number of ionizing photons available from the AGN in that same year to within ${\sim}$ 10\%\footnotemark \footnotetext{independent of the gas density and the filling factor.} which is considered to be good, possibly fortuitous, agreement given the large extinction correction and the known UV and X-ray variability \citep{Mar09, Vas09}. Nevertheless, the comparison suggests that the central AGN is apparently able to ionize the entire BLR in NGC 3227 which necessarily entails a high,  ${\sim}$ 100\%,  covering factor. Even if 44\% of the broad H${\alpha}$ emission line flux results from collisional excitation, rather than photoionization as alluded
to in section 2.0.2, the covering factor is still high ${\sim}$ 62\%. Such a high covering factor is easy to achieve with a spherical distribution of gas and impossible to achieve with a thin 
accretion disk, further favoring an inflow as the origin of the broad Balmer emission lines seen in NGC 3227.

\subsection{BLR Size}

\cite{Kas05} noted that NGC 3227 does not conform to the
correlation between reverberation size and UV luminosity established for quasars and high
luminosity AGNs. However, \cite{Cre01} present evidence for considerable dust extinction to the central UV source in NGC 3227 amounting to ${\sim}$ 4.3 mag at 1450{\AA}. The consequence
of matching the extinction corrected 1450{\AA} UV luminosity, estimated using Figure 1 to be 1.68 ${\times}$ 10${^{42}}$ erg s${^{-1}}$, with the smaller reverberation size measured by \cite{Den09, Den10} is to translate NGC 3227 closer to the correlation defined by quasars and high luminosity AGNs thereby reducing the discrepancy first noted by \cite{Kas05}.

Adopting a BH mass allows the gravitational radii, constrained using the inflow model (Section 3.3), to be converted into a physical size.
If one adopts the reverberation BH mass of 7.6 ${\times}$ 10${^6}$ M$_{\odot}$ \citep{Den09, Den10}, then the inner and outer radii of the model inflow correspond to ${r_i}$ ${\sim}$ (3.2 ${\pm}$ 0.6) l.d and ${r_o}$ ${\sim}$ (89 ${\pm}$ 36) l.d, respectively.
Thus, the inner radius of the model inflow consistently coincides with the reverberation radius (3.8 ${\pm}$ 0.8 l.d.) reported by \cite{Den09,Den10}. The most straightforward interpretation of this result is that reverberation mapping locates just the inner radius of a much larger volume of ionized gas. 
Presumably, reverberation mapping locates the inner radius because this is where the number density of inflowing gas filaments is highest and the gas density in the filaments is also highest as explained in Section 4.4, but the emission is inevitably diminished {\it in a spectrum} to zero intensity for smaller radii by an ever increasing Doppler spread. Thus, the inner radius is actually an illusion. Since the emitting region is ionization bounded, the notion of an outer radius is also an illusion. Consequently, what we perceive as the BLR in NGC 3227, visualized in Figure 7,  is just the illuminated shell of a much larger volume of inflowing gas whose actual dimensions remain to be determined. A measure of the inflow rate can be judged from the animation presented in Figure 7 which shows that the inflowing gas takes about 28 weeks to fall from the reverberation radius into the BH.

\subsection{Photoionization Model}

The physical dimensions of the ionized region in NGC 3227 are consistent with the predictions of a photoionization model analogous to the one described for HII regions in \cite{Ost06}. In the model, the number of photoionizations is balanced by the number of recombinations at each location in the nebula according to

\begin{equation}
\frac {n_H^o(r) \int^{\nu_{max}} _{\nu_{T}} N({\nu})_{ion} a(\nu)e^{-\tau(\nu,r)}d{\nu}}{ 4 \pi r^2} = n_e(r)^2 \alpha_{B}
\end{equation}

where ${n_H^o}$ is the number density of neutral H atoms, ${n_e}$ is the number density of electrons (equal to the number of protons in a pure H nebula), and ${N(\nu)_{ion}}$ is the number of ionizing photons per unit frequency, ${\nu}$, characterized previously in section 4.1. 
The photoionization cross-section, ${a(\nu)}$, presented in \cite{Ost06}, is modified to accommodate the consequence of the Doppler effect in diminishing ${a(\nu)}$ for infalling H atoms. Specifically, the photoionization cross section for inflowing hydrogen atoms is blue shifted with respect to the central AGN and hence diminished by the strong,  ${\nu^{-3}}$, dependence of the photoionization cross section on frequency. The easiest way to visualize this is from the perspective of an inflowing H atom, for which the 13.6 eV ionization edge of the central continuum UV-X-ray source blueshifts to higher frequencies as the H atom accelerates towards the BH.  
The optical depth, ${\tau(\nu,r)}$, is computed using

\begin{equation}
\tau(\nu,r) = \int^{r} _{0} n_H^o(r) a(\nu) dr
\end{equation}

and is then substituted into equation 2 and solved for the electron density, ${n_e(r)}$. The integral within each radius is implemented numerically by adopting a distribution ${n_H^o(r) \propto r^{-3/2}}$, identical to that
employed to describe the inflow (see section 3.2). The integral over frequency is from the ionization threshold, ${\nu_{T}}$, 
(which depends on inflow velocity according to the relativistic Doppler equation) to ${\nu_{max}}$ = 3.29 ${\times}$ 10$^{19}$ Hz. The calculation yields the radial dependence of the electron density, ${n_e(r)}$, the ionization fraction, ${n_e(r)}$/(${n_e(r)}$ + ${n_H^o(r)}$), and the ionization parameter, ${\Gamma(r)}$, given by 

\begin{equation}
\Gamma(r) = N_{ion}/ 4  \pi r^2 c~ n_e
\end{equation}

Figure 8 illustrates the radial dependence of these three quantities resulting from photoionization of the neutral gas, ${n_H^o(r)}$.  
Both the electron density (${n_e(r) \propto r^{-5/2}}$) and the ionization fraction ( ${\propto r^{-1/2}}$) increase 
rapidly ${\it towards}$ the central AGN leading to the visual impression of a compact BLR. Most of the inflowing gas is completely ionized by the time it reaches the reverberation radius. On the other hand the ionization parameter 
(${\Gamma(r) \propto r^{1/2}}$)  increases rapidly ${ \it away}$ from the central AGN to achieve values that are several orders of magnitude larger than those typically employed for BLRs in the literature \citep[e.g.,][and references therein]{Kwa84, Fer92, Lew03, Era10}. Values cited in the literature for the ionization parameter are derived, typically,
from emission line ratios using photoionization codes \citep[e.g., {\it Cloudy},][]{Fer98, Col88, Lei07}.  Thus, there appears to be a large discrepancy between values quoted in the literature for the ionization parameter in BLRs and those implied by the photoionization model presented here. A similar discrepancy has been noted previously by \cite{Ale94}.

\subsection{Virial Black Hole Masses}

The virial product of reverberation radius, ${R}$,  and the velocity width\footnotemark 
\footnotetext{the FWHM or the second moment \citep{Col06}.} of the broad Balmer line, ${\Delta}V$, leads to a BH mass according to the relation;

\begin{equation}
M_{\bullet} = f R {\Delta}V^{2}/G
\end{equation}

where ${\it f}$ is a correction factor that locates the reverberation BH masses on the  M$_{\bullet}$--${\sigma}_* $ relation defined by BH masses measured directly using gas and stellar kinematics. ${\it f}$ has a value of ${\sim}$ 5.5 \citep{Onk04}. However, if the broad Balmer lines {\it are} produced by an inflow then it follows that the reverberation radius corresponds to the inner radius of the flow. Figure 7 visualizes the model inflow and identifies 8\% of the broad H${\beta}$ line flux that is observed to be time variable\footnotemark \footnotetext{The H${\alpha}$ variability amplitude is observed to be ${\sim}$ 11\%
with STIS. However, adopting 11\% of the flux as the time variable component does not change the visual impression depicted in Figure 8 that reverberation mapping locates just the inner edge of a much larger volume of gas.} \citep{Win95, Den09, Den10}. Figure 9 illustrates that the full width at zero intensity of the broad emission line profile is defined by the inner radius of the inflow. 
Consequently, the correct virial product, at least for NGC 3227, is the reverberation radius (0.0035 pc) and the half width at zero intensity (4300 km/s) squared. Then one recovers the reverberation BH mass of 7.6 ${\times}$ 10${^6}$ M$_{\odot}$ with ${\it f}$ = 1. Thus, there is no need for the correction factor, ${\it f}$, if the virial product utilizes a radius with the maximum velocity measured ${\it at~that~same~radius}$.

\subsection{Physical Properties of the Inflow}

In the context of AGNs, gas inflows have received remarkably little attention in the published literature which is ironic because 
virtually everyone agrees that AGN activity is fueled by inflowing gas. Single peak broad emission line profiles,
which are by far the most common \citep{Str03}, may in particular prove to be a lucrative resource for quantifying the physical properties of gas inflows by providing perhaps the only
way to measure a number of key parameters including the mass of emitting gas, the ionized gas volume filling factor and the mass inflow rate. Specific results for NGC 3227 are summarized in Table 5, following the procedure described most recently in \cite{Dev11b}. An important result is that the observed mass inflow rate, ${\sim 10^{-2}~\rm M_\odot~yr^{-1}}$, is commensurate with that required to power the central UV-X-ray source
in NGC 3227, assuming radiatively inefficient accretion as parameterized by \cite{Mer03}. 

Two outstanding conceptual problems remain with regard to fueling AGN. The first is why the accretion rate implied by the bolometric luminosity of the AGN is so low compared to the available gas supply \cite[e.g.,][]{Ho09}. The second unsolved problem is how does the gas lose angular momentum in order to be accreted \citep[e.g.,][and references therein]{Beg94, Pro03}. Both of these difficulties can be solved simultaneously if the source of the inflowing material is zero angular momentum gas resulting from stellar-mass loss. Only a small fraction of the stellar mass loss gas is expected to have zero angular momentum but this naturally explains why the accretion rate is observed to be low ${\sim 10^{-2}~\rm M_\odot~yr^{-1}}$. Not only is the accretion rate implied by the bolometric luminosity of the AGN low, the accretion rate is {\it actually observed to be low} and the explanation offered here for that observation is that the BH accretes {\it only the fraction of gas produced by stellar mass loss that has zero angular momentum}. A corollary of this explanation is that inflowing gas has no rotation; gas with any appreciable angular momentum will not
migrate inward for the reasons outlined by \cite{Beg94}. Thus, inflowing gas, by definition, has no rotation and consequently no angular momentum.

The [S II] lines have a line width corresponding to 360 km/s indicating that they arise from gas
at a radial distance of ${\sim}$0.5 pc. Thus, the two [S II]  lines provide an opportunity to measure the electron density well beyond the region of the inflow that is ionized by the central AGN. 
The nominal value for the observed [S II] ${\lambda}$6742/${\lambda}$6757 
intensity ratio = 0.78 ${\pm}$ 0.05, corresponding to ${n_e}$ = 10$^3$ cm$^{-3}$. The fact that the FWHM of the broad Balmer  emission lines is larger than that of the [O I] and [O III] emission lines (Table 2 \& 3) suggests that the electron density ${\it inside}$ the inflow must be greater than the critical density of the levels from which the  [O I] and [O III] lines originate which is ${\sim}$10$^6$ cm$^{-3}$. This limit is higher than the one obtained from the [S II] line ratio implying that the electron density increases towards the central BH, consistent with the radial dependence predicted by photoionization model (section 4.3).

The vacuum wavelength [O III] ${\lambda}$4960.30 and ${\lambda}$5008.24 lines are observed to have a FWHM ${\sim}$800 km/s,
corresponding to a radius of ${\sim}$0.1 pc, providing the opportunity to measure the electron temperature just beyond the outer boundary of the inflow that is ionized by the central AGN. The limit for the observed [O III] ratio (${\lambda}$4960.30 + ${\lambda}$5008.24) / ${\lambda}$4364.44 ${\geq}$ 56 yields T ${\leq}$ 16,650 K for the electron temperature if ${n_e}$ = 10$^3$ cm$^{-3}$.

Both \cite{Den09} and \cite{Win11} have raised the possibility of outflowing gas ${\it at}$ and ${\it inside}$\footnotemark \footnotetext{Based on the ionization parameter measured
for the O VII and O VIII absorption lines by \cite{Win11}} the reverberation radius in NGC 3227. 
However, the low Eddington ratio dictates that radiation pressure acting on the ionized gas fails to overcome the gravitational force
by a factor of ${\sim}$ 250. Furthermore, a dust driven outflow \citep{Fab06} seems unlikely as the dust reverberation radius \citep[${\sim}$20 l.d.,][]{Sug06} lies well beyond the gas reverberation radius \citep[${\sim}$4 l.d.,][]{Den09} as illustrated in Figure 7. Additionally, any advection driven outflow is predicted to occur on a size scale comparable to $r_g$ \citep{Nar97} which is ${\sim}$1000 times smaller than the gas reverberation radius in NGC 3227. Thus, it is incumbent on proponents of outflow to identify a viable mechanism that operates over 4${\pi}$ steradians and is able to thwart the ballistic pressure of the inflow, estimated to be ${\sim}$ 0.4 erg/cm$^3$ at the gas reverberation radius and increasing as $r^{-2}$ towards the central BH. 

\subsection{Summary}

Collectively, the results presented here for NGC 3227 challenge many preconceived notions concerning the nature of BLRs in active galactic nuclei. Firstly, there are quite likely no {\it broad line clouds} but rather ionized gas {\it filaments}. Secondly, the BLR is {\it not small}, it is just {\it perceived} to be small. Thirdly, the single peak broad Balmer emission lines, which are by far the most common type of broad emission line seen in AGNs, are quite likely the signature of a sub-parsec sized {\it ballistic 
inflow}, not a Bondi flow. Fourthly, the observed mass inflow rate {\it is} commensurate with that required to power the AGN in terms of radiatively inefficient accretion. However, it remains to be determined to what extent these findings 
for NGC 3227 apply to more luminous AGNs. Nevertheless, the picture emerging from analysis of {\it HST}/STIS observations, to date, is that the BLR of LLAGNs are very different from each other with no evidence to support the notion of a {\it standard, classical} or {\it normal} BLR.

\section{Conclusions}

Spectroscopic observations of NGC 3227 with the {\it Hubble Space Telescope} have revealed a time variable single-peak broad H${\alpha}$ emission line profile 
which has been successfully modeled as a steady state inflow. The {\it perceived} dimensions of the BLR correspond to 
an outer radius of ${\sim}$90 l.d. and an inner radius of ${\sim}$3 l.d. that coincides with the gas reverberation radius. However, the {\it perceived} small size for the BLR is an illusion, a consequence
of the fact that the emitting region is ionization bounded at the outer radius and diminished by Doppler broadening at the inner radius. The actual dimensions of the inflow are likely much larger
and remain to be determined. If the electron density is high, ${n_e}$ ${\geq}$ 10$^6$ cm$^{-3}$, as suggested by the absence of similarly broad 
[O I] and [O III] emission lines, then the luminosity in the broad H${\alpha}$ emission line observed in the year 2000 leads to a steady state inflow rate ${\sim} 10^{-2} ~{\rm M_\odot~yr^{-1}}$ which is sufficient to explain the X-ray luminosity of the AGN, if the AGN is powered by radiatively inefficient accretion.




\acknowledgments
This research has made extensive use of the NASA Astrophysics Data System, the Atomic Line List, http://www.pa.uky.edu/~peter/newpage/
and a variety ${\bf STSDAS}$ tasks. The author would like to thank Emily Heaton her assistance and the anonymous referee for a thoughtful and charitable report. 



{\it Facilities:}  \facility{HST (STIS)}

\clearpage



\begin{figure}
\epsscale{1.0}
\begin{center}
\plotone{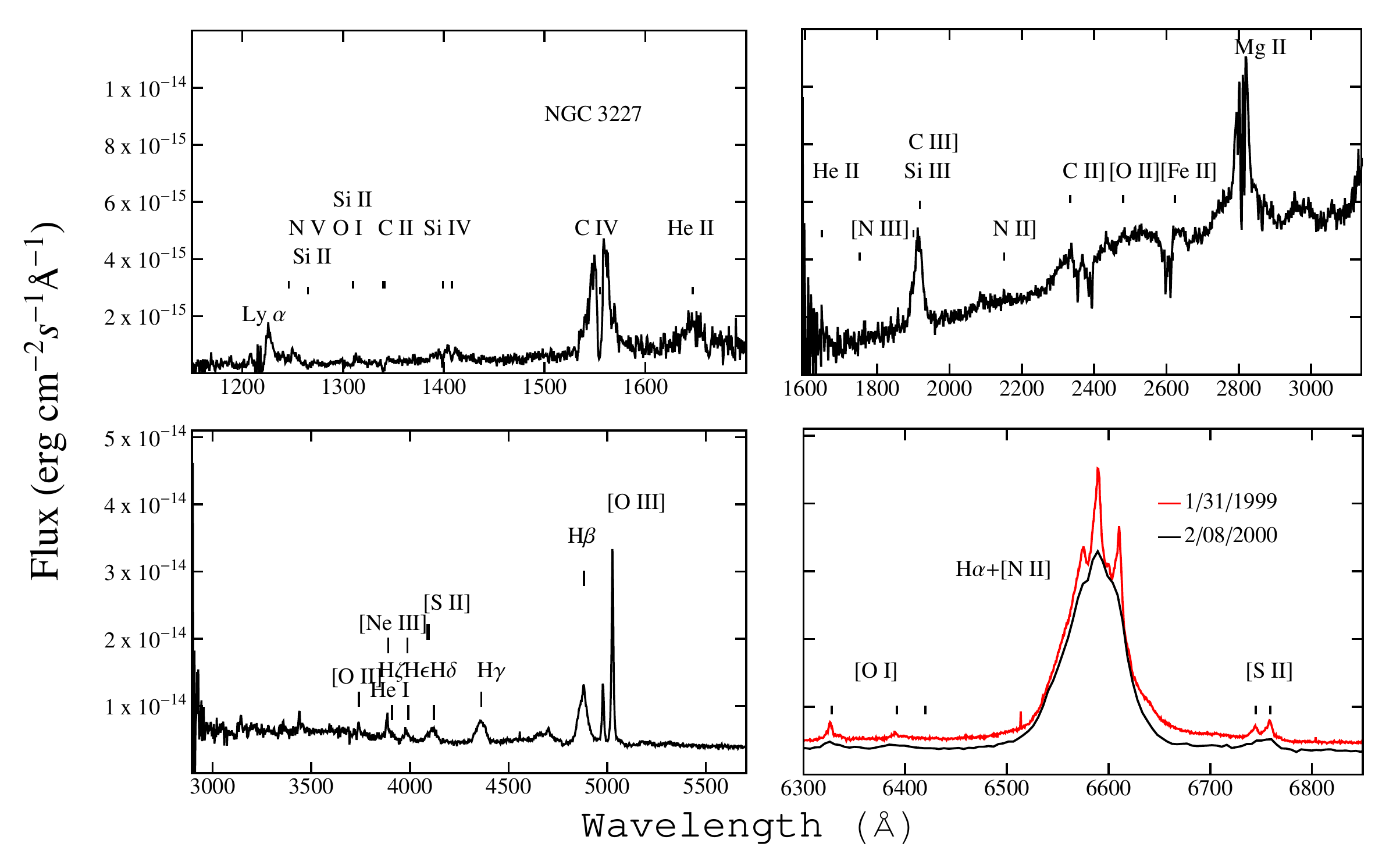}
\caption{{ Visual and UV spectra  of NGC 3227 as seen through the following gratings:  {\sl Top left panel}: G140L. {\sl Top right panel}: G230L.
{\sl Lower left panel}: G430L.  {\sl Lower right panel}: G750M. Red line shows data obtained under PID 7403. Black lines for all panels show data obtained under PID 8479.}}
\label{default}
\end{center}
\end{figure}

\clearpage

\begin{figure}
\epsscale{1.0}
\begin{center}
\plotone{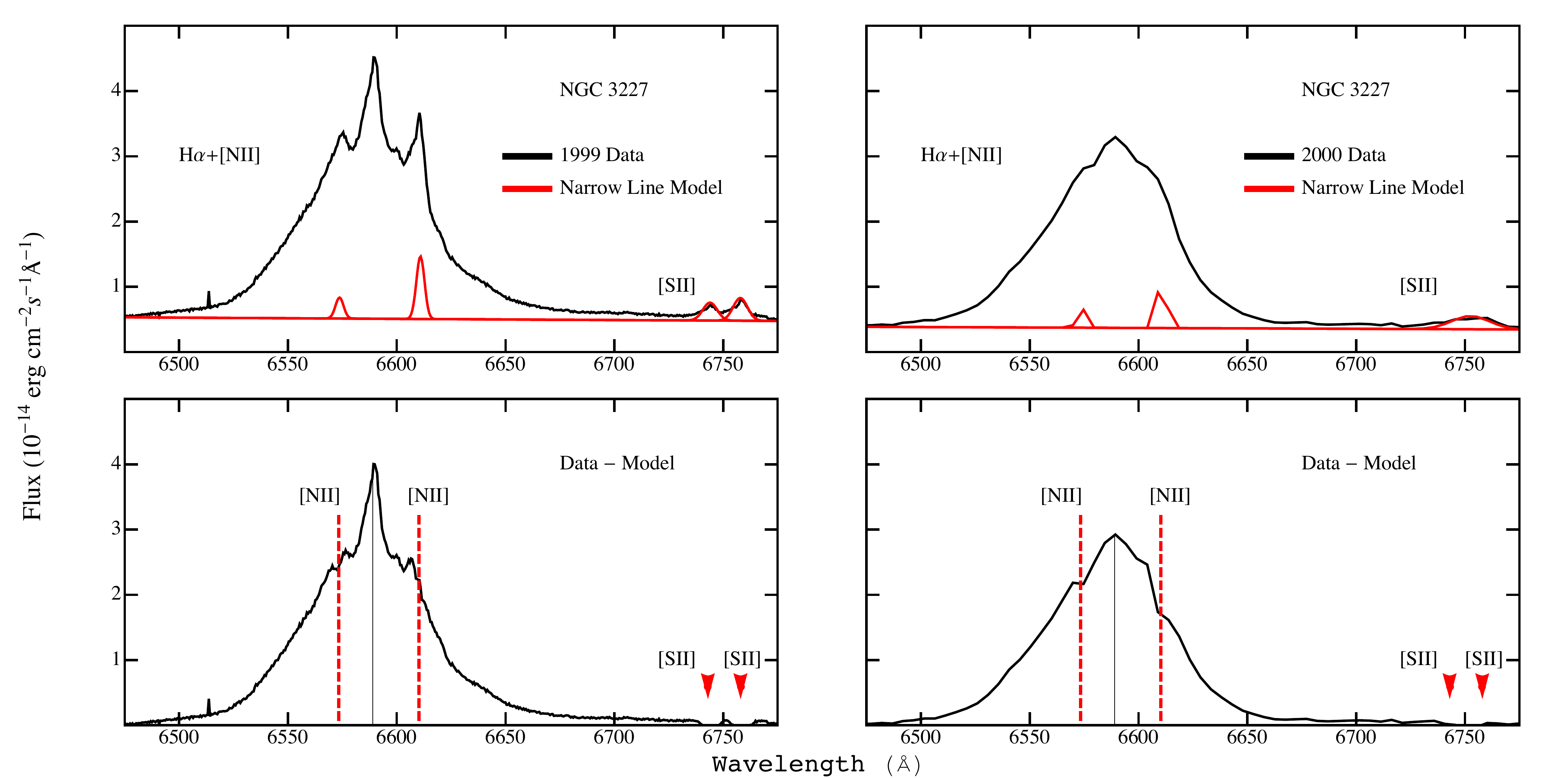}
\caption{{ Broad H${\alpha}$ emission line in NGC 3227 observed in 1999 with G750M ({\sl Left panels}) and 2000 with G750L ({\sl Right panels}). {\sl Top panels}: The observed spectrum is shown in black and a model for the forbidden lines is shown in red (see also Table 2). {\sl Lower panels}: The broad 
H${\alpha}$ emission line profile
after the forbidden lines have been subtracted. The central wavelengths of the subtracted lines are
indicated in red. The vertical black line corresponds to the observed (redshifted) central wavelength of the H${\alpha}$ line }}
\label{default}
\end{center}
\end{figure}

\clearpage

\begin{figure}
\epsscale{0.8}
\begin{center}
\plotone{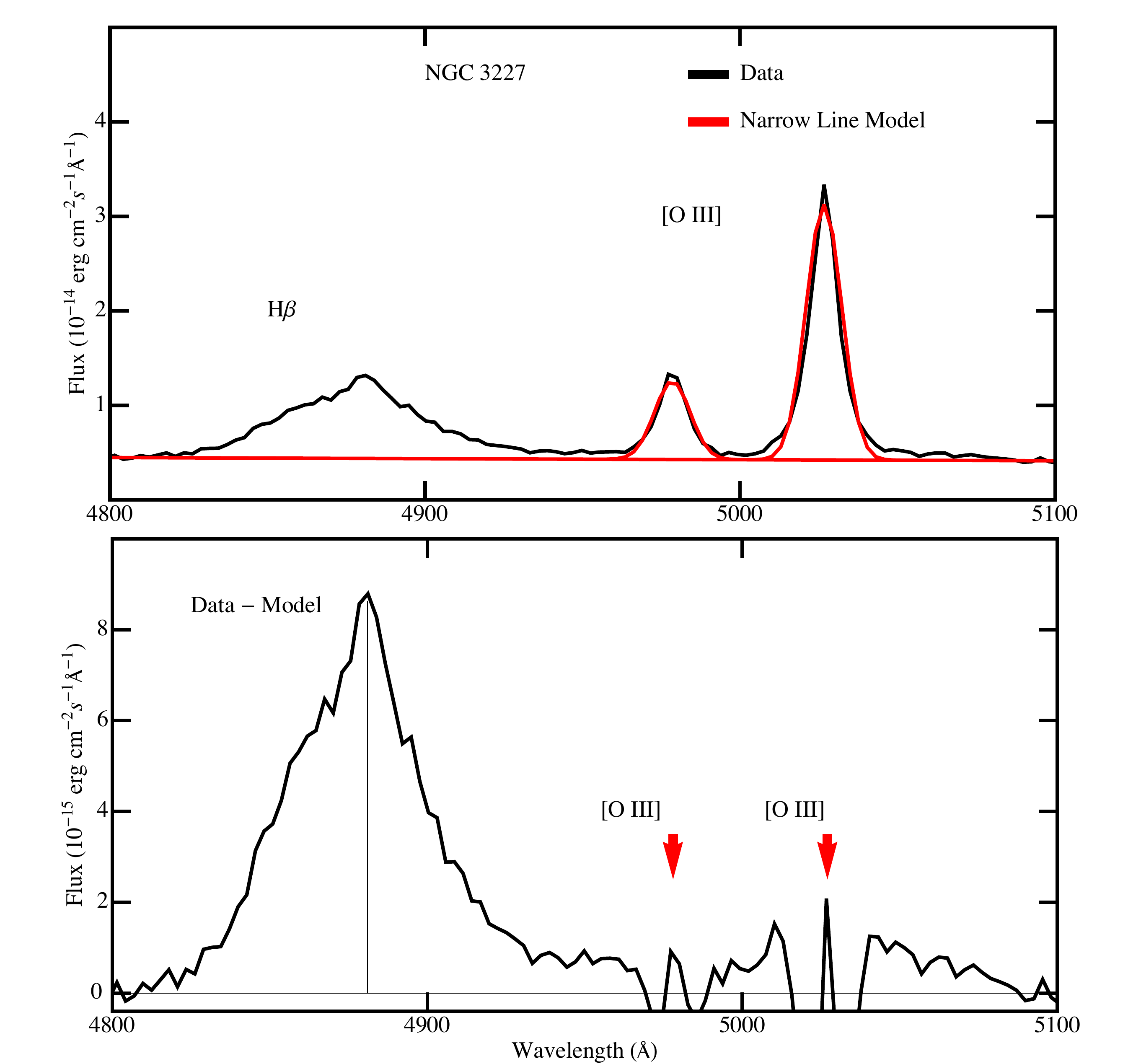}
\caption{{ Broad H${\beta}$ emission line in NGC 3227 observed in 2000. {\sl Top panel}: The observed spectrum
is shown in black and a model for the forbidden lines is shown in red (see also Table 3). {\sl Lower panel}: The broad 
H${\beta}$ emission line profile after the forbidden lines have been subtracted. The central wavelengths of the subtracted lines are
indicated in red. The vertical black line corresponds to the observed (redshifted) central wavelength of the H${\beta}$ line }}
\label{default}
\end{center}
\end{figure}

\clearpage

\begin{figure}
\epsscale{0.8}
\begin{center}
\plotone{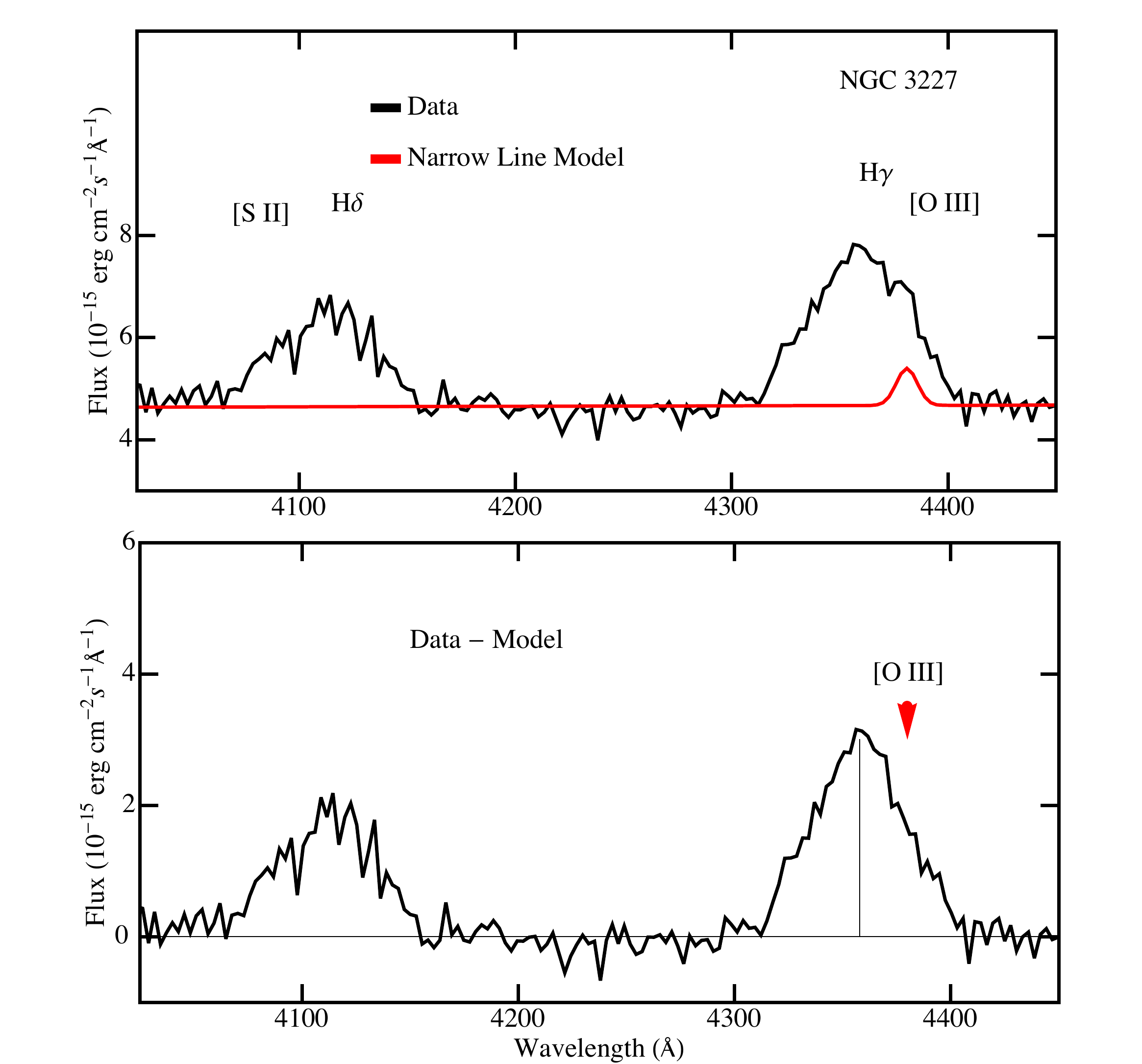}
\caption{{ Broad H${\gamma}$ emission line in NGC 3227 observed in 2000. {\sl Top panel}: The observed spectrum
is shown in black and a model for the forbidden line is shown in red (see also Table 3). {\sl Lower panel}: The broad 
H${\gamma}$ emission line profile
after the forbidden line has been subtracted. The central wavelength of the subtracted line is
indicated in red. The vertical black line corresponds to the observed (redshifted) central wavelength of the H${\gamma}$ line }}
\label{default}
\end{center}
\end{figure}

\clearpage

\begin{figure}
\epsscale{1.0}
\begin{center}
\plotone{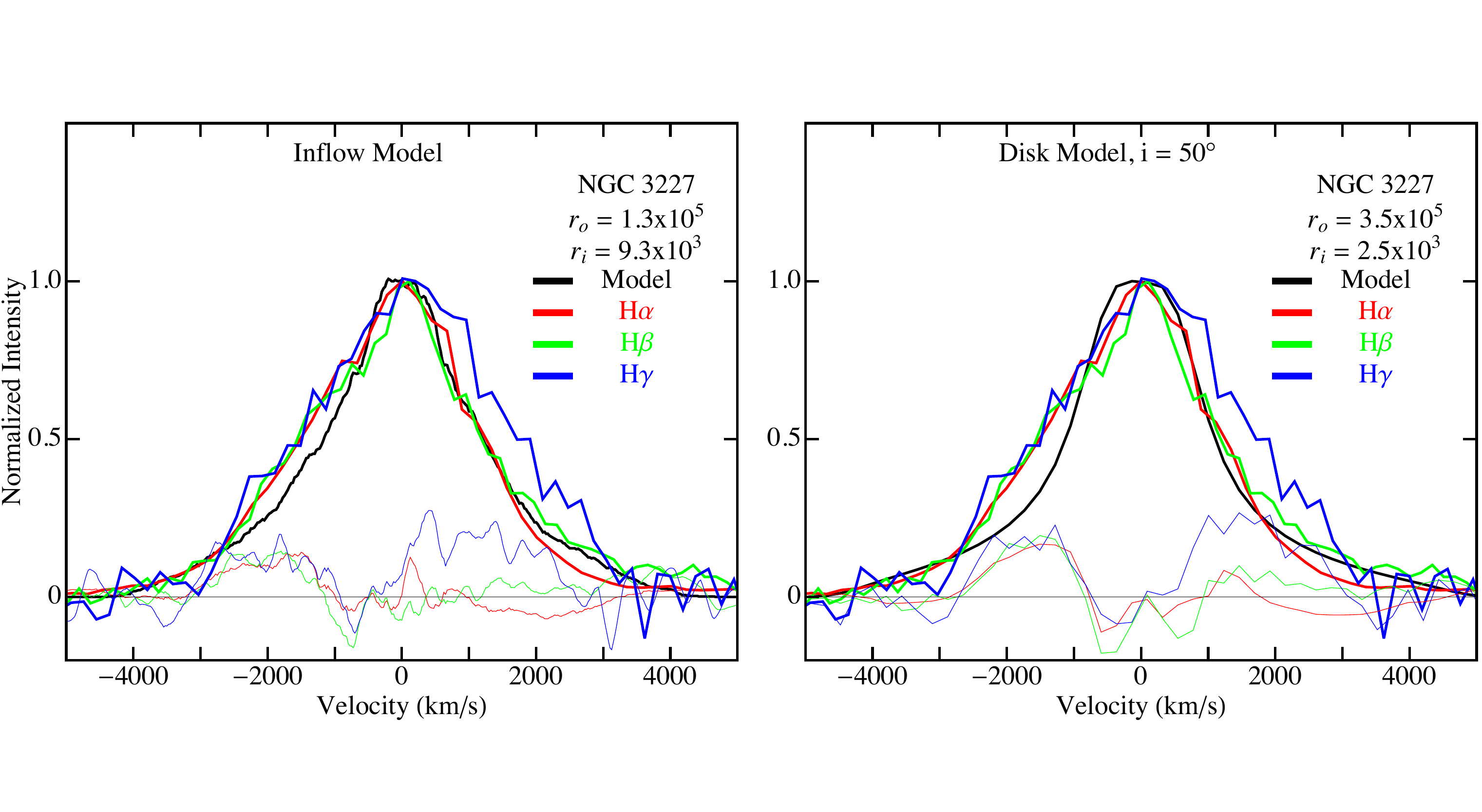}
\caption{{Model representation of the broad Balmer lines observed with the G750L grating (Left panel)  in terms of a spherically symmetric inflow and (Right panel) in terms of a relativistic accretion disk. The 
observed H${\alpha}$, H${\beta}$ and H${\gamma}$ emission lines are shown in red, green and blue, respectively. The inflow and disk models are shown in black. The inner, $r{_i}$, and outer radii, $r{_o}$, are indicated in units of gravitational radii, $r{_g}$. Residuals for each profile are plotted as thinner colored lines.}}
\label{default}  
\end{center}
\end{figure}

\clearpage

\begin{figure}
\epsscale{1.0}
\begin{center}
\plottwo{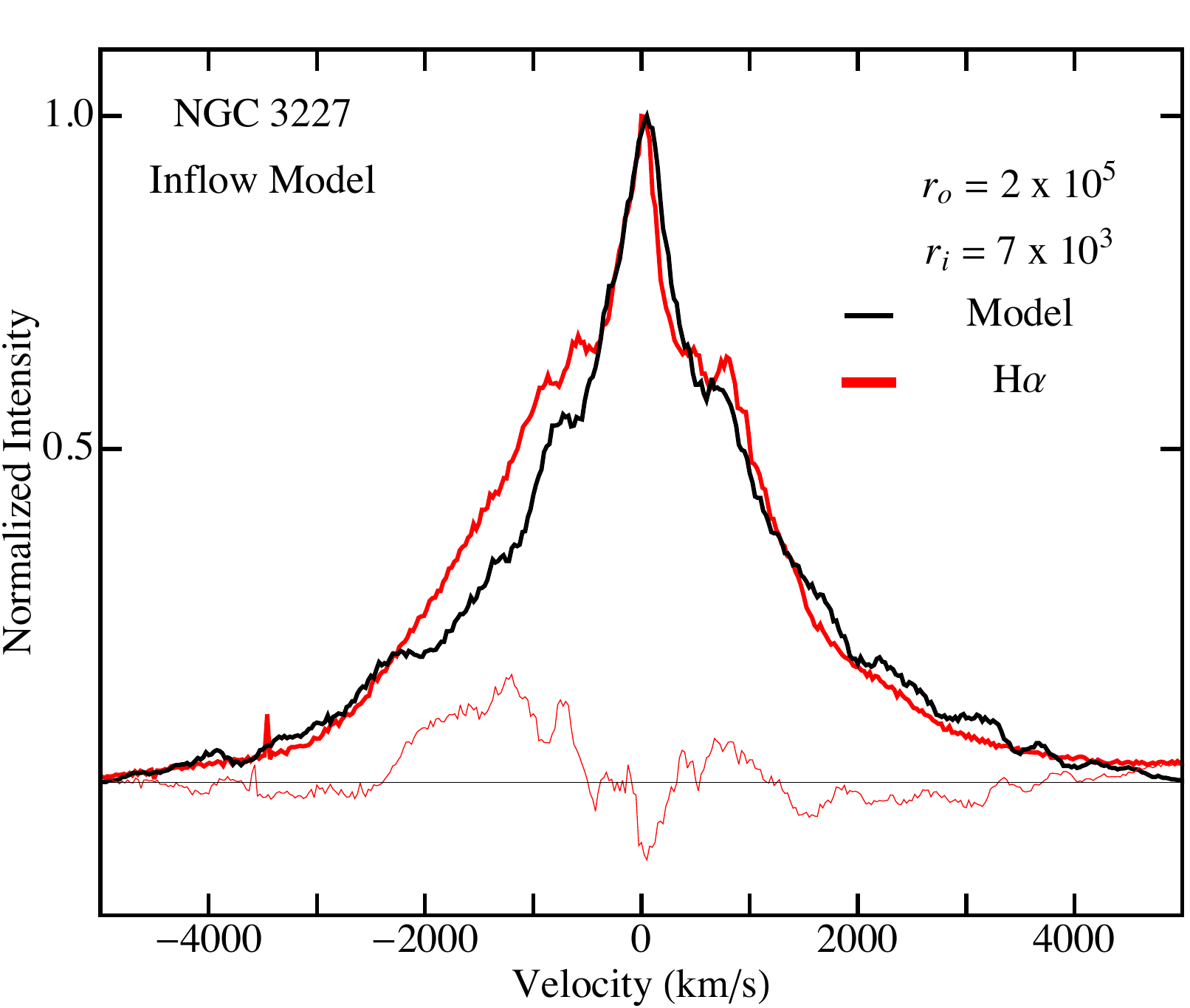}{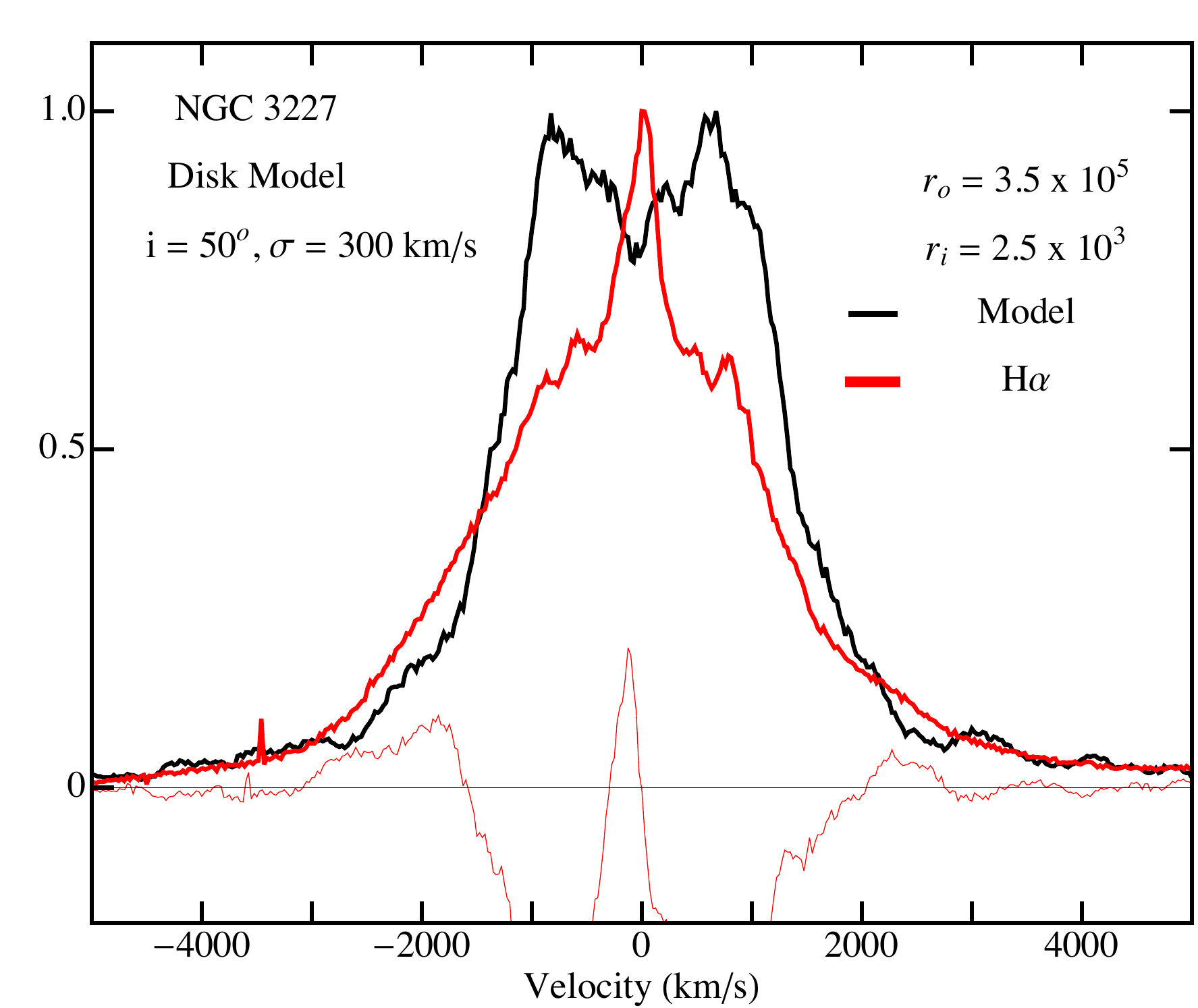}
\caption{{Model representation of the broad H${\alpha}$ line observed with the G750M grating (Left panel) in terms of a spherically symmetric inflow and (Right panel) in terms of an accretion disk. The 
observed H${\alpha}$ emission line is shown in red. The inflow and disk models are shown in black. The inner, $r{_i}$, and outer radii, $r{_o}$, are indicated in units of gravitational radii, $r{_g}$. Residuals for each profile are plotted as thinner red lines.}}
\label{default}  
\end{center}
\end{figure}

\clearpage

\begin{figure}
\epsscale{0.8}
\begin{center}
\plotone{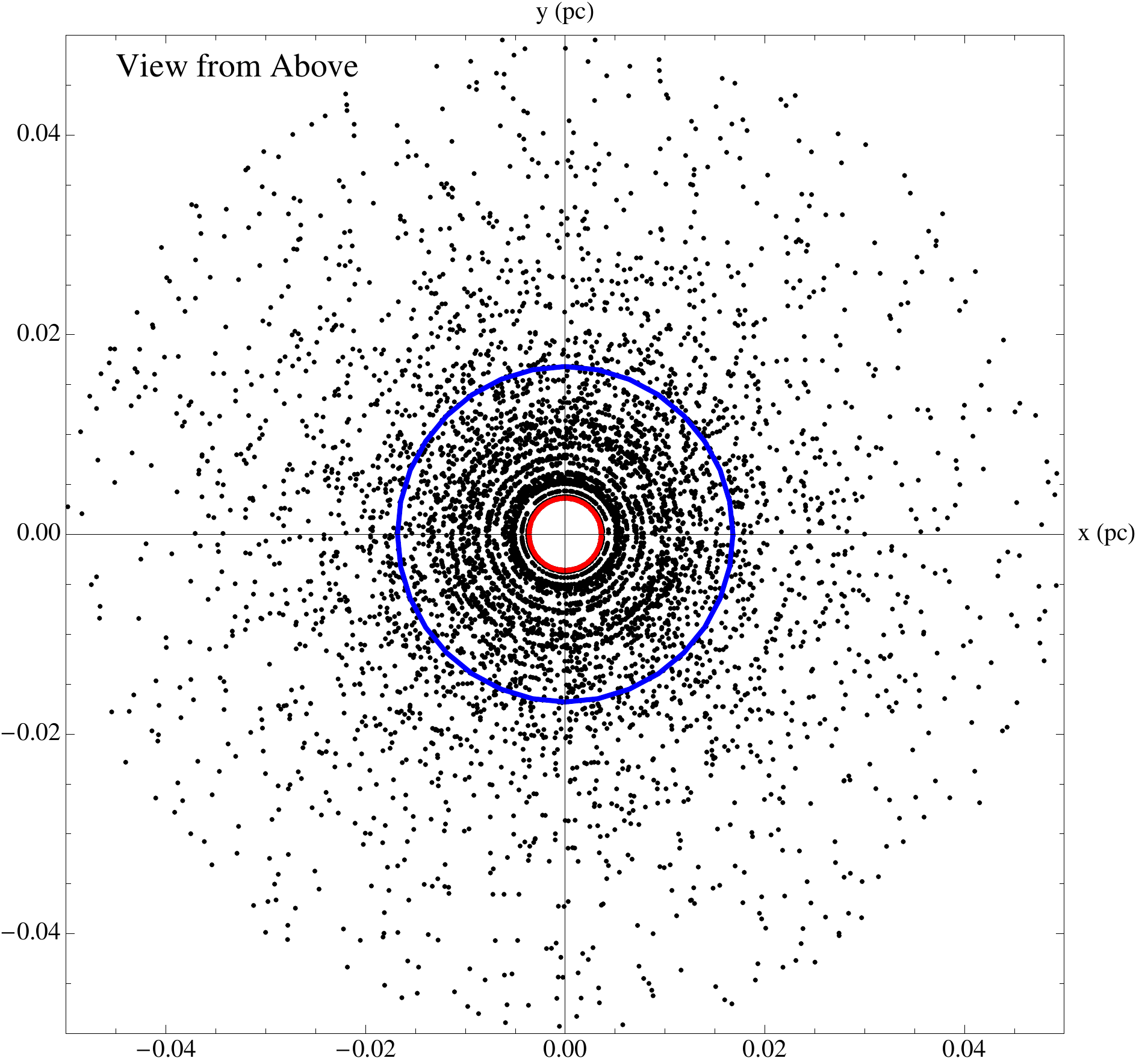}
\caption{{{ Visualization of the model inflow. The red ring identifies 8\% of the points representing the percentage of the total broad line flux which is observed to respond to time-variable illumination from the central AGN \citep{Den09,Den10}. The blue ring identifies the $K$--band reverberation radius  \citep{Sug06}. This figure is the first frame in an animation that is accessible in the electronic edition of the {\it Astrophysical Journal}. }}}
\label{default}
\end{center}
\end{figure}

\clearpage

\begin{figure}
\epsscale{1.0}
\begin{center}
\plotone{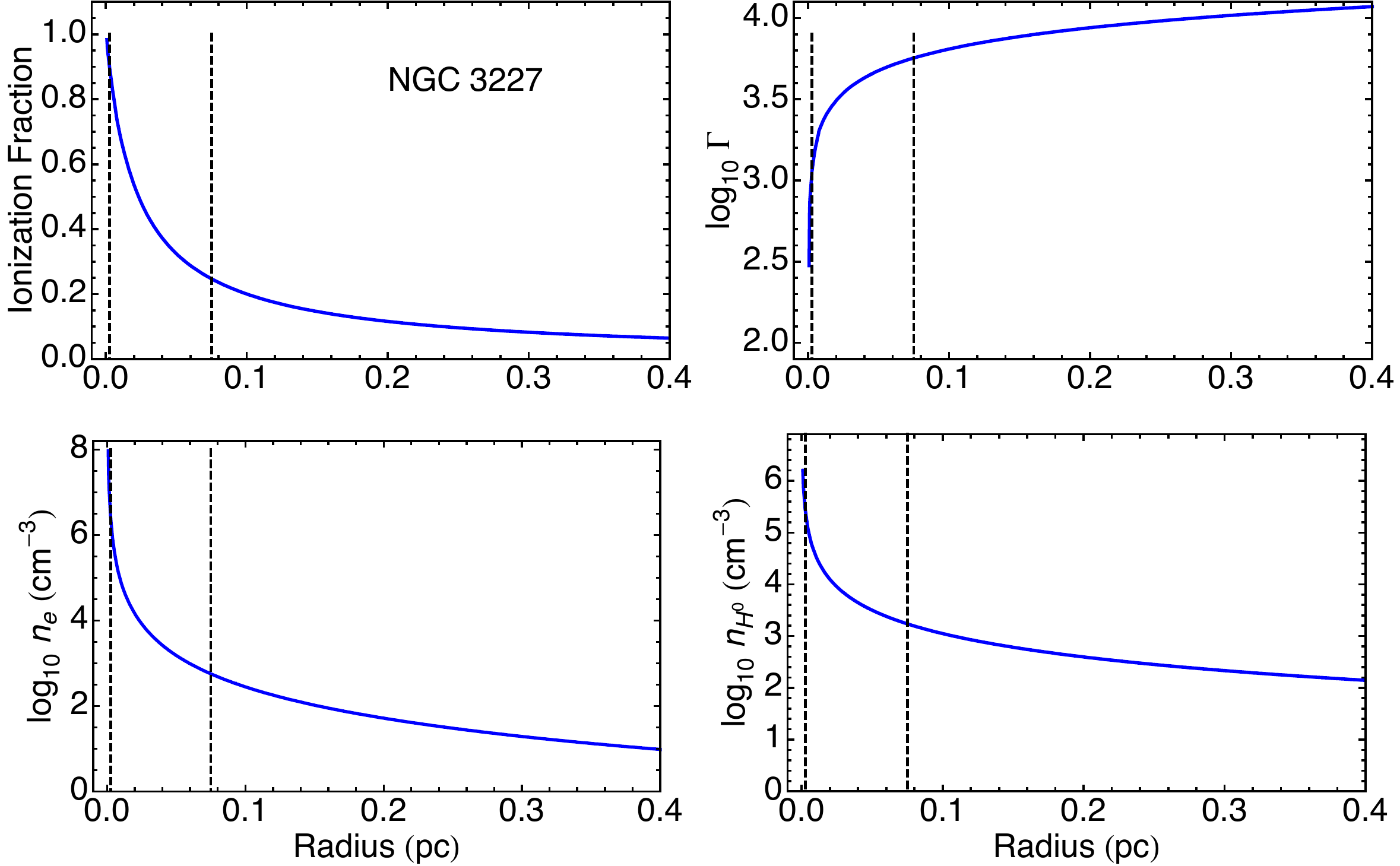}
\caption{{Photoionization model results for NGC 3227 illustrating the radial dependence of the ionization fraction, ${\propto r^{-1/2}}$ (Top left panel), the ionization parameter, ${\propto r^{1/2}}$
(Top right panel), the electron number density, ${\propto r^{-5/2}}$ (Lower left panel) and the neutral H number density, ${\propto r^{-3/2}}$ (Lower right panel). Vertical dashed lines identify the inner, $r{_i}$, and outer radii, $r{_o}$, of the BLR in units of pc (see section 4.2). }}
\label{default}  
\end{center}
\end{figure}

\clearpage

\begin{figure}
\epsscale{0.7}
\begin{center}
\plotone{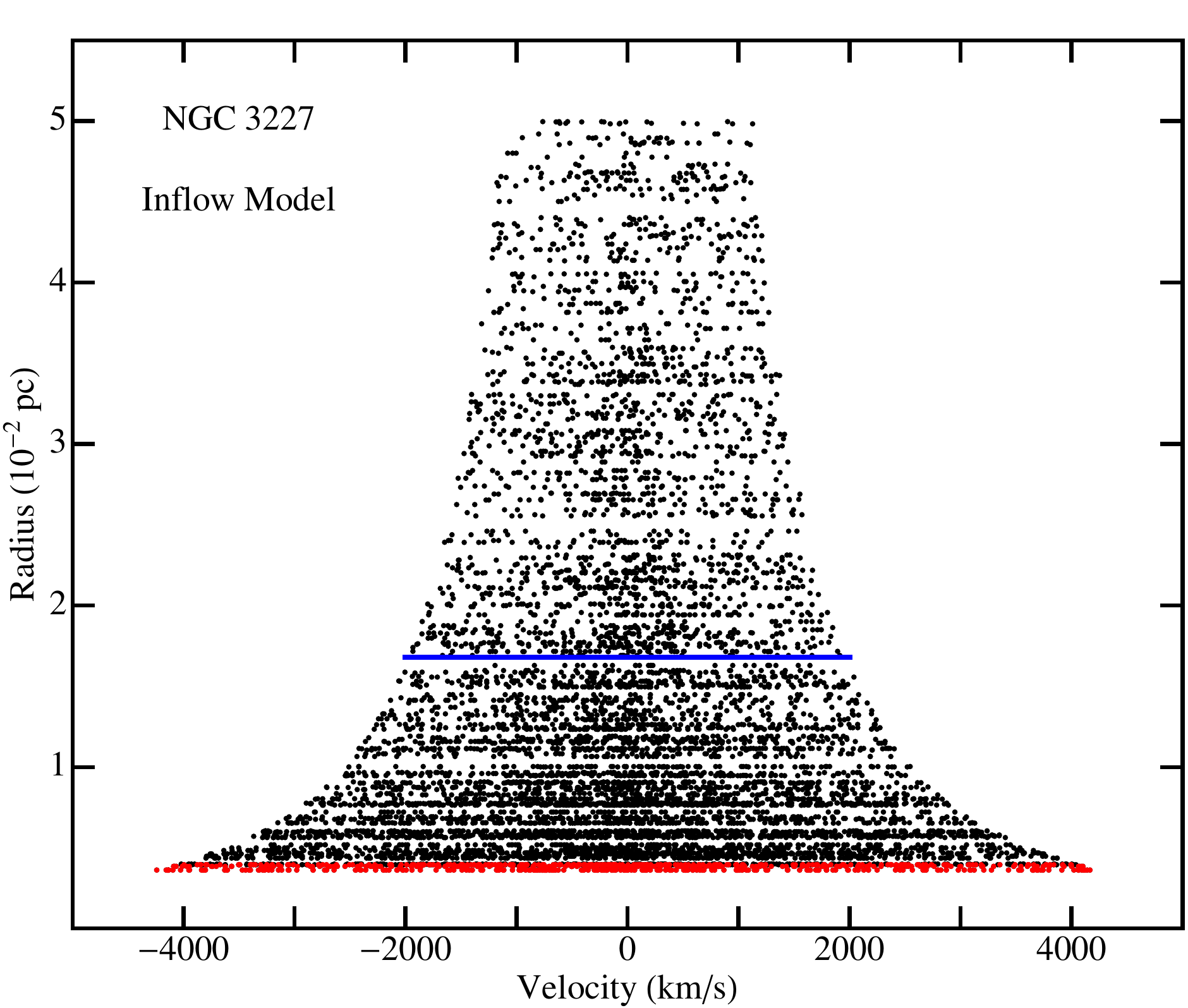}
\caption{{ The number density of points in the model inflow. Red line identifies 8\% of the points representing the percentage of the total broad line flux which is observed to respond to time-variable illumination from the central AGN \citep{Den09,Den10}. The blue line identifies the $K$-band reverberation radius \citep{Sug06}. The corresponding model broad H${\alpha}$ profiles, shown in Figures 5 and 6, are produced by summing the points vertically. }}
\label{default}
\end{center}
\end{figure}

\clearpage

\begin{deluxetable}{ccccccccc}
\tabletypesize{\scriptsize}
\tablecaption{NGC 3227 Spectral Datasets\label{tbl-2}}
\tablewidth{0pt}
\tablehead{
\colhead{PID} & \colhead{Observation Date} & \colhead{Grating} & \colhead{Spectral Range} & \colhead{Slit} & \colhead{Dispersion} & \colhead{Plate Scale} & \colhead{Integration Time} & \colhead{Datasets}   \\
\colhead{} & \colhead{} &  \colhead{} & \colhead{\AA}  & \colhead{arc sec} & \colhead{\AA/pixel}& \colhead{arc sec/pixel} & \colhead{s} & \colhead{}\\
\colhead{(1)} & \colhead{(2)} &  \colhead{(3)} & \colhead{(4)}  & \colhead{(5)} & \colhead{(6)} & \colhead{(7)} & \colhead{(8)} & \colhead{(9)} \\
}
\startdata
7403 & 1-31-1999 &  G750M  & 6295 -- 6867 & 52 x 0.2 & 0.56 & 0.05 & 1890 & o57204040 \\
8479 & 2-8-2000 &  G230L  & 1570 -- 3180 & 52 x 0.2 & 1.58 & 0.025 & 1612 & o5kp01010 \\
8479 & 2-8-2000 & G430L & 2900 -- 5700 & 52 x 0.2 & 2.73 & 0.05 & 120 & o5kp01020 \\
8479 & 2-8-2000 & G140L & 1150 -- 1736 & 52 x 0.2 & 0.6 & 0.024 & 2132 & o5kp01030 \\
8479 & 2-8-2000 & G750L & 5236 -- 10266 & 52 x 0.2 & 4.92 & 0.05 & 120 & o5kp01040 \\
\enddata

\end{deluxetable}

\clearpage

\begin{deluxetable}{cccc}
\tabletypesize{\scriptsize}
\tablecaption{Emission Line Parameters for the G750M Nuclear Spectrum Obtained 1--31--1999\tablenotemark{a}}
\tablewidth{0pt}
\tablehead{
\colhead{Line} & \colhead{Central Wavelength\tablenotemark{b}} & \colhead{Flux\tablenotemark{c}} & \colhead{FWHM}   \\
\colhead{} & \colhead{\AA} &  \colhead{10$^{-15}$ erg cm$^{-2}$ s$^{-1}$} & \colhead{kms$^{-1}$} \\
\colhead{(1)} & \colhead{(2)} &  \colhead{(3)} & \colhead{(4)} \\
}
\startdata
$\textrm{[O I]}$ &  6326  ${\pm}$ 1 & 12.9 ${\pm}$ 0.7 &  298 ${\pm}$  24 \\
$\textrm{[O I]}$  &  6390  ${\pm}$ 1 & 6.3 ${\pm}$ 0.9 &  298 ${\pm}$  52 \\
$\textrm{[N II]}$ &  6574 ${\pm}$ 3   & 15  &  200   \\
H${\alpha}$ (broad)\tablenotemark{d} & 6588  & 2200 ${\pm}$ 1 & 2147 ${\pm}$ 100 \\
H${\alpha}$ (broad)\tablenotemark{e} & 6588  & 1950 ${\pm}$ 3 & 2700 ${\pm}$ 200 \\
$\textrm{[N II]}$ & 6610 ${\pm}$ 1  & 45\tablenotemark{f} & 200   \\
$\textrm{[S II]}$ & 6744 ${\pm}$ 1  & 20.8 ${\pm}$ 2\tablenotemark{g}  & 357 ${\pm}$ 47 \\
$\textrm{[S II]}$ & 6758 ${\pm}$ 1 & 26.6  ${\pm}$ 2\tablenotemark{g}  & 357 ${\pm}$ 34 \\
\enddata
\tablenotetext{a}{Table entries that do not include uncertainties are fixed parameters.
}
\tablenotetext{b}{Observed wavelength}
\tablenotetext{c}{Measured within
a 0.2{\arcsec}  x 0.35{\arcsec}  aperture. Continuum subtracted but not corrected for dust extinction. Model dependent systematic uncertainties introduce an additional ${\sim}$3\% error not reported in the Table.}
\tablenotetext{d}{Observation date: 1--31--1999, G750M}
\tablenotetext{e}{Observation date: 2--8--2000, G750L}
\tablenotetext{f} {The [N II] emission line flux is chosen so as to not over-subtract the broad H${\alpha}$ emission line profile. }
\tablenotetext{g} {The [S II] emission lines are unresolved in the lower resolution spectrum obtained with the G750L grating on 2--8--2000. However,
the sum of the flux in the two lines then was (43.7 ${\pm}$ 6) $\times$10$^{-15}$ erg cm$^{-2}$ s$^{-1}$ which concurs with the sum of the flux in the 
[S II] lines measured on 1--31--1999 reported in the Table.}
\end{deluxetable}

\clearpage

\begin{deluxetable}{cccc}
\tabletypesize{\scriptsize}
\tablecaption{Emission Line Parameters for the G430L Nuclear Spectrum Obtained 2--8--2000\tablenotemark{a}}
\tablewidth{0pt}
\tablehead{
\colhead{Line} & \colhead{Central Wavelength\tablenotemark{b}} & \colhead{Flux\tablenotemark{c}} & \colhead{FWHM}   \\
\colhead{} & \colhead{\AA} &  \colhead{10$^{-14}$ erg cm$^{-2}$ s$^{-1}$} & \colhead{kms$^{-1}$} \\
\colhead{(1)} & \colhead{(2)} &  \colhead{(3)} & \colhead{(4)} \\
}
\startdata
$\textrm{[O II]}$ &  3737  ${\pm}$ 1 &  ${\leq}$ 0.6 &  ... \\
H${\gamma}$ (broad) & 4356 &  15.6 ${\pm}$ 0.4 & 3600 ${\pm}$ 300 \\
$\textrm{[O III]}$\tablenotemark{d} &  4381  &  ${\leq}$ 0.9  &  778 \\
H${\beta}$ (broad) & 4882  &  49.2 ${\pm}$ 0.01 & 3000 ${\pm}$ 300 \\
$\textrm{[O III]}$ &  4978  ${\pm}$ 1 & 12 ${\pm}$ 1 &  801 ${\pm}$ 131 \\
$\textrm{[O III]}$ &  5027  ${\pm}$ 1 & 38 ${\pm}$ 2 &  788 ${\pm}$  49 \\

\enddata
\tablenotetext{a}{Table entries that do not include uncertainties are fixed parameters. }
\tablenotetext{b}{Observed wavelength}
\tablenotetext{c}{Measured within
a 0.2{\arcsec}  x 0.35{\arcsec}  aperture. Continuum subtracted but not corrected for dust extinction. Model dependent systematic uncertainties introduce an additional ${\sim}$3\% error not reported in the Table.}
\tablenotetext{d} {The [O III] emission line parameters chosen so as to not over-subtract the broad H${\gamma}$ emission line profile}

\end{deluxetable}

\clearpage

\begin{deluxetable}{cccc}
\tabletypesize{\scriptsize}
\tablecaption{Stellar Mass Model}
\tablewidth{0pt}
\tablehead{
\colhead{r} & \colhead{r} & \colhead{M(r)\tablenotemark{a}} & \colhead{$\rho$(r)\tablenotemark{b}}   \\
\colhead{pc} & \colhead{mas} &  \colhead{10$^{6}$ M$_\odot$} & \colhead{10$^{6}$M$_\odot$/pc$^3$} \\
\colhead{(1)} & \colhead{(2)} &  \colhead{(3)} & \colhead{(4)} \\
}
\startdata
1 & 9.90 &  5.56  & 1.3 \\
0.5 &  4.90  &  1.63  &  3.1 \\
0.25 & 2.47 &  0.48 & 7.3 \\
0.125 &  1.24   & 0.14  &  16.5 \\
0.0625 &  0.62  & 0.04 &  36.2 \\

\enddata
\tablenotetext{a}{Enclosed stellar mass calculated by integrating the $H$-band surface brightness profile parameterized by \cite{Hic08}, adopting an $H$-band mass-to-light ratio M/L = 0.7 and +3.65 for the absolute $H$-band magnitude for the Sun \citep{Dev87}. }
\tablenotetext{b}{Stellar mass density}

\end{deluxetable}

\clearpage

\begin{deluxetable}{cccc}
\tabletypesize{\scriptsize}
\tablecaption{Physical Properties of the Inflow}
\tablewidth{0pt}
\tablehead{
\colhead{Parameter} & \colhead{Value} \\
\colhead{(1)} & \colhead{(2)}  \\
}
\startdata
Electron number density, ${\it n{_e}}$ & ${\ge}$  10$^6$ cm$^{-3}$ \\ 
H${{\alpha}}$ luminosity\tablenotemark{a}, $L (H{{\alpha}}$) & 2.4 x 10${^7}$ L${_{\sun}}$ \\
Mass of ionized gas in BLR, $M_{emitting}$ & ${\le}$ 222 M${_{\sun}}$ \\
Volume filling factor of ionized gas in BLR, ${ \epsilon}$ & ${\sim}$ 1  \\
Ionized gas mass inflow rate, ${\dot{m}}$ & ${\sim 1.2 \times 10^{-2} ~\rm M_\odot~yr^{-1}}$ \\
2-10 keV X-ray luminosity\tablenotemark{b}, $L_{2-10~keV}$ & $3.2\times 10^{41}~{\rm erg~s^{-1}}$ \\
Mass inflow rate required by AGN, ${\dot{m}}$ & ${\sim 1.8 \times 10^{-2} ~\rm M_\odot~yr^{-1}}$ \\

\enddata
\tablenotetext{a}{measured in the year 2000 (Table 2)}
\tablenotetext{b}{measured in the year 2000 \citep{Vas09}}

\end{deluxetable}

\end{document}